\documentclass[conference]{IEEEtran}

\usepackage[usenames,dvipsnames]{xcolor}
\usepackage[english]{babel}

\usepackage{amsmath}
\usepackage{amsthm}
\usepackage{graphicx}
\usepackage{algorithm}
\usepackage{algorithmic}
\usepackage{amssymb}
\usepackage{bm}
\usepackage{lipsum}

\usepackage[binary-units=true]{siunitx}

\def\R{{\mathbf R}}
\def\N{{\mathbf N}}
\def\Z{{\mathbf Z}}
\def\E{{\mathbf E}}

\def\1{{\mathbf 1}}
\newtheorem{theorem}{Theorem}
\newtheorem{lemma}{Lemma}

\newtheorem{assumption}{Assumption}
\newtheorem{definition}{Definition}

\DeclareMathOperator*{\argmin}{\arg \min}
\DeclareSIUnit{\bits}{bits}
\DeclareSIUnit{\bit}{bit}
\DeclareSIUnit{\byte}{B}
\DeclareSIUnit{\cycles}{cycles}
\DeclareSIUnit{\flops}{FLOPS}
\DeclareSIUnit{\frame}{frame}

\title{Maximum Lifetime Analytics in IoT Networks}

\author{\IEEEauthorblockN{V\'ictor Valls\IEEEauthorrefmark{1}, George Iosifidis\IEEEauthorrefmark{1}, Theodoros Salonidis\IEEEauthorrefmark{2}}\\
	\IEEEauthorblockA{\IEEEauthorrefmark{1}School of Computer Science and Statistics, Trinity College Dublin, Ireland}
	\IEEEauthorrefmark{2}IBM T.J.\ Watson Research Center, Yorktown Heights, NY, USA
	\thanks{{This work was supported by Science Foundation Ireland under Grant No. 17/CDA/4760.}}%
\vspace{-0.5em}
}

\IEEEoverridecommandlockouts
\begin{document}
\maketitle

\begin{abstract}
This paper studies the problem of allocating bandwidth and computation resources to data analytics tasks in Internet of Things (IoT) networks. IoT nodes are powered by batteries, can process (some of) the data locally, and the quality grade or performance of how data analytics tasks are carried out depends on {where} these are executed. The goal is to design a resource allocation algorithm that jointly maximizes the network lifetime and the performance of the data analytics tasks subject to energy constraints. 
This joint maximization problem is challenging with coupled resource constraints that induce non-convexity.  We first show that the problem can be mapped to an equivalent convex problem, and then  propose an online algorithm that provably solves the problem and does not require any a priori  knowledge of the time-varying wireless link capacities and data analytics arrival process statistics. The algorithm's optimality properties are derived using an analysis which, to the best of our knowledge, proves for the first time the convergence of the dual subgradient method with time-varying sets.
Our simulations seeded by real IoT device energy measurements, show that the network connectivity 
plays a crucial role in network lifetime maximization, that the algorithm can obtain both maximum network lifetime and maximum data analytics performance in addition to maximizing the joint objective, and that the algorithm increases the network lifetime by approximately 50\% compared 
to an algorithm that minimizes the total energy consumption.
%
\end{abstract}

\section{Introduction}\label{sec:introduction}

We consider an Internet of Things (IoT) network where a set of nodes collect measurements that have to be later on analyzed by a data analytics or machine learning (ML) algorithm. For example, an algorithm for classifying, filtering or summarizing data. 
This class of services is one of the most important envisioned applications of the emerging IoT networks \cite{linda-ICC17} and poses many technical challenges \cite{EU-IoT}; especially, when IoT networks operate subject to {bandwidth}, {processing}, and {energy} constraints.

Unlike previous generations of sensor networks, it is expected that IoT applications collect data at an unprecedented rate and that only a fraction of these will be non-ephemeral \cite{cisc-cloud-index,idc-whitepaper}. 
Hence, the usual approach of transferring all the data to an IoT node gateway for processing, even if possible, may consume network resources unnecessarily. A promising solution to overcome this issue is to leverage the processing capacity of the IoT nodes and execute (some of) the data analytics tasks at the edge. That is, by carrying out some of the processing in situ it is possible to reduce the amount of data that needs to be transmitted over the network.
However, executing data analytics at the IoT nodes has some costs. The two most significant ones are perhaps that processing is expensive in terms of energy, and that the grade or performance of {how} a task is carried out depends on the algorithms that IoT nodes are able to run. For example, nodes with limited memory can only classify data with low complexity models, which makes their predictions less accurate in general.

In this paper, we introduce a resource allocation framework that allows us to design online execution and routing policies for data analytics tasks in IoT networks. That is, decide whether  a data analytics task should be processed (i) locally, (ii) at the gateway, or (iii) elsewhere in the IoT network (\emph{e.g.}, at a neighboring IoT node) depending on the available network resources  (bandwidth, processing capacity and energy). An important feature of the framework is that it also allows us to maximize a combined criterion of  \emph{network lifetime}\footnote{Time the network can operate without any node running out of battery.} and \emph{data analytics performance}. The first objective is important because we would like the network to operate for as long as possible, and the second because the network has to fulfill its purpose besides ``staying alive''.  We call this problem Maximum Lifetime IoT Analytics (MLIA).

The problem of maximizing the time a network can operate has been addressed before in sensor networks (see Section \ref{sec:related_work}). However, unlike sensor networks, IoT encompasses more sophisticated scenarios where a variety of heterogeneous devices and applications coexist \cite{EU-IoT}, and brings data analytics processing into play. The latter adds, technically, a new dimension to the existing routing algorithms (\emph{e.g.}, \cite{CT04,CT99, ML06, XCN05}) and raises technical challenges that cannot be addressed  with previous solutions directly. 
In particular, to capture in-network processing we need to use a network model with ``gains'' (Section \ref{sec:vcgains}), transform a non-convex problem into an equivalent convex one (Lemma \ref{th:maxlifeconvex}),  and develop a new online algorithm (Theorem \ref{th:main_theorem}) that can handle non-linearities in the utility as well as randomness in the actions (\emph{i.e.}, the routing and processing decisions that can be made in each time slot).  Furthermore, the fact that nodes can only operate during a limited time span\footnote{Due to energy constraints.} adds the difficulty of selecting the algorithm parameters so as to obtain the desired performance (see Section \ref{sec:performance}). 
To this end, the main contributions of the paper are:

\begin{itemize}
\item [(i)] \underline{\smash{MLIA problem}}: we introduce the problem of jointly maximizing the lifetime of an IoT network and the performance of how data analytics tasks are carried out. This is an open problem arising in many IoT applications.

\item [(ii)] \underline{\smash{Problem model}}: we formulate the MLIA problem as a convex optimization program. The model captures aspects such as the coexistence of different types of data analytics tasks, that data analytics may be carried out by different IoT nodes, and that their computation cost may vary across  nodes, among others. 

\item [(iii)] \underline{\smash{Online algorithm}}: we propose an online algorithm that solves  the underlying convex problem and has non-asymptotic convergence guarantees.  The algorithm determines the joint routing and processing policy for the IoT nodes in a myopic manner by only looking at the system's current state, \emph{i.e.}, it does not require statistical knowledge of the underlying random processes such as the time-varying link capacities. Also, and to the best of our knowledge, this is the first paper that presents the convergence of the dual subgradient method with time-varying sets, which is another contribution. 


\item [(iv)] \underline{\smash{Performance evaluation}}: we perform a set of extensive experiments to (i) evaluate the performance of the proposed solution and to (ii) understand how this is affected by the network connectivity and  system parameters. We also compare our algorithm to two benchmark policies, and show that our approach can increase the network lifetime by approximately $50 \%$ compared to an algorithm that minimizes the total energy consumption. 
\end{itemize}

The rest of the paper is organized as follows. Section \ref{sec:related_work} presents the related work. In Section \ref{sec:model}, we introduce the system model, the problem, and the arising trade-offs. In Section \ref{sec:maxlife}, we formulate the MLIA problem as a convex program, and in Section \ref{sec:dynamic_problem}, we  present the online algorithm. Section \ref{sec:evaluation} contains the numerical experiments and discussion.  

\section{Related Work}\label{sec:related}
\label{sec:related_work}

The problem of deciding how to transmit and process data to prolong the network lifetime has been studied before in Wireless Sensor Networks (WSNs). For instance, the work in \cite{ZS09} proposes load-balancing techniques to spread the energy consumption across nodes. In \cite{CZ05}, the authors study the problem of designing a medium access protocol that takes into account the channel state information and the available energy. And in \cite{CTV06}, it is shown that preprocessing data at the sensor nodes can help to reduce the network load and so the energy required to transmit data to the {fusion center} (\emph{i.e.}, the gateway). From a problem formulation perspective, our approach differs from previous works in the literature of WSNs and IoT because we consider both the energy spent in routing and processing the data, and the performance of the analytics. The last point is crucial in heterogenous IoT networks where the ability of nodes to run algorithms depends on their hardware.

The maximum lifetime objective has been extensively studied for routing in multi-hop sensor networks. The seminal work in \cite{CT99} considered a static maximum lifetime routing problem and formulated it as a linear program. This work was extended in a sequel of papers \cite{CT00, CT04} to different types of wireless networks, and a \emph{flow augmentation} algorithm was proposed to support fixed and arbitrary generation rates. The approach in \cite{CT99} has also been adopted by other authors (see survey \cite{CS16}); for example, \cite{ZCB+07} combines network lifetime with congestion control, and \cite{ML06} proposes an algorithm to solve the maximum lifetime problem in a distributed fashion. Finally, \cite{GIS18} proposed a static optimization solution for maximizing analytics performance in IoT networks with average power constraints, which is different than the lifetime criterion.
Our work differs technically from all previous work on maximum lifetime routing in sensor networks because (i) it considers IoT sensing nodes that are heterogeneous and can perform data analytics computations in addition to routing data; (ii) it uses an objective that incorporates performance in addition to lifetime; and (iii) the proposed dynamic algorithm solves, provably, the underlying convex problem without knowledge of the arrivals or channel statistics. 
Regarding the algorithm, we use time-varying sets to handle the instantaneous routing and processing constraints, which is in marked contrast to stochastic Lagrange dual approaches where the stochasticity does not affect the decision variables.

Finally, we note that data analytics optimization with routing costs has been considered in the cloud context \cite{FLT+16, FLT+17}.
However, there the costs are not related to energy expenditure, neither the resource allocation decisions affect the time the network will be able to operate. Furthermore, in IoT, we have the additional inherent difficulties of routing data over wireless networks which do not appear, of course, in the cloud.


\section{System Model and Problem Statement} \label{sec:model}

\subsection{System model}

\subsubsection{Network} We model an IoT network as a directed graph $\mathcal G=(\mathcal{N}, \mathcal{E})$ consisting of $n = | \mathcal{N} |$ nodes and $l = | \mathcal{E} |$ links. We use $\mathcal{C}$ to denote the set of applications in the network. For each application $c \in \mathcal{C}$, the nodes collect and send data to a gateway node---possibly over a multi-hop path.  Parameter $\lambda_i^{(c)}$, $i \in \mathcal N$, $c \in \mathcal C$ indicates the rate at which node $i$ collects data of application $c$. These are exogenously given and injected into the network directly. The arrival rate matrix is given by
\begin{align}
\bm{\lambda}=\big(\lambda_{i}^{(c)}: i \in \mathcal{N}, c \in \mathcal{C} \big). \label{eq:arrival_matrix}
\end{align}

Each link $(i,j)\in\mathcal{E}$ has an average capacity of $\mu_{ij}$ \si{\bits/\second}; we collect these in matrix $\bm{\mu} = (\mu_{ij} : i,j \in \mathcal{N})$. Links $(i,j)$ and $(j,i)$ can have different capacities.  
The transmission over link $(i,j)$ induces an energy consumption $e_{ij}^{\text{tx}}$ for the sender node $i$, and $e_{ij}^{\text{rx}}$ for the receiver node $j$. Both are measured in \si{\joule/\bit}. We model the wireless interference using the protocol interference model \cite{kumar-interference}, according to which a transmission over link $(i,j)\in\mathcal{E}$ is successful if and only if all nodes in range with $i$ or $j$ are idle. This requirement is based on the CSMA/CA protocol adopted by IEEE 802.11 standards.\footnote{In detail, this interference model complies with the communication sequence RTS-CTS-Data-ACK, where each sender is also a receiver of the ACK packets; which is the strictest model. The set of interfering nodes can be reduced if the ACK operation is not used.}
The set of interfering links for each link $(i,j)\in\mathcal{E}$ is defined as: 
\begin{align}
I(i,j):=\{ (a,d), (d,b): d\in\mathcal{N}_i\cup\mathcal{N}_j, a, b \in\mathcal{N}_d  \},
\label{eq:I}
\end{align}
where $\mathcal{N}_i:=\{ j\in\mathcal{N}: (i,j)\in\mathcal{E} \}$ is the set that contains the neighbors of a node $i \in \mathcal N$.
Hence, when a link $(i,j)$ is active none of the links in $I(i,j)$ can be used.

\subsubsection{Nodes} IoT nodes may be heterogenous in terms of hardware. We use $\rho_i$ to denote the processing capacity in \si{\flops} of a node $i \in \mathcal N$, and $\gamma_{i}^{(c)}$ to denote the processing requirements (in \si{\flops/\bit}) of an application in each of the nodes.  The processing in each node may reduce the volume of each flow by a factor of $0 \le \beta_{i}^{(c)} \le 1$.  For example, in an object recognition application the flow volume reduction is large since an image gets reduced to a collection of bounding boxes and tags \cite{RF18}, \emph{i.e.}, to few bytes. Of course, this flow reduction factor may depend on where a data analytics task is carried out since IoT nodes may run different algorithms. 
We collect the flow reduction and processing requirements in matrices $\bm{\beta}=(\beta_{i}^{(c)}:i\in\mathcal{N}, c\in\mathcal{C} )$ and $ \bm{\gamma}=(\gamma^{(c)}_{i}: i \in \mathcal N, c\in\cal C )$.

Each IoT node $i \in \mathcal N$ has an energy budget of $E_i$ Joules that can spend transmitting, receiving, and processing data. When a node has used its energy budget, it dies meaning that it cannot transmit or process more data. We assume that the network gateway does not have energy constraints.  

\subsection{Decision variables and constraints}
\label{sec:vcgains}

%
IoT nodes can make two types of decisions: process and forward data. Variable $x_{ij}^{(c)}\geq 0$ indicates the rate in \si{\bits/\second} at which application $c$ is transmitted over link $(i,j)$. Similarly, variable $y_{i}^{(c)}$ indicates the rate in \si{\bits/\second} that node $i$ processes data of application $c$. We collect the decision variables in matrices
$\bm{x}  = (x_{ij}^{(c)}:\,(i,j)\in\mathcal{E},\,c\in\mathcal{C})$ and $\bm{y}  = ( y_{i}^{(c)}:\,i\in\mathcal{N},\,c\in\mathcal{C} )$.
The transmission and processing rates must satisfy the link and node capacity constraints: 
\begin{align}
 \sum_{c \in \mathcal C} x^{(c)}_{ij} \le \mu_{ij} ,  \quad \sum_{c \in \mathcal C} \gamma_{i}^{(c)} y^{(c)}_{i} \le \rho_i \quad \forall i \in \mathcal N,  (i,j) \in \mathcal E.  \label{eq:capacity_constraints}
\end{align}
%
The interference constraints affect how the links in $\mathcal{E}$ can be activated and consequenctly the total amount of data that can be transferred over the network; see, for instance, Lemma 1 in \cite{kodialam-mobicom}. These are formally given by:
\begin{align}
	\sum_{c\in\mathcal{C}} \frac{x_{ij}^{(c)}}{\mu_{ij}}+\sum_{(k,m)\in I(i,j) }\sum_{c\in\mathcal{C}}\frac{x_{km}^{(c)}}{\mu_{km}}\leq 1 && \forall\,(i,j)\in\mathcal{E},\label{eq:interference_constraint}
\end{align}
where $I(i,j)$ is defined in (\ref{eq:I}).

\begin{figure}
	\centering
	\includegraphics[width=\columnwidth]{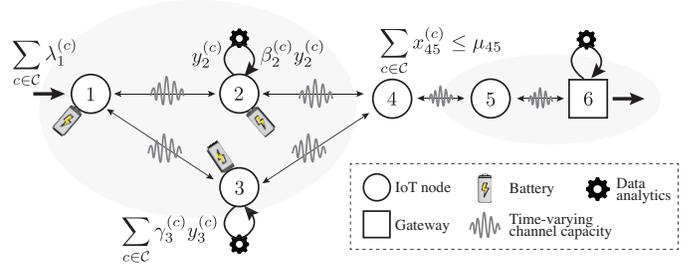}
	\caption{Illustrating an example of an IoT network with energy constraints. Node 1 receives data analytics tasks that must be processed either at node 2, node 3, or at the gateway. Nodes 1, 2 and 3 have energy constraints and the transmission of data over the network is affected by the interference (large shaded areas) and link capacity constraints. }
	\label{fig:proc_network}
	\vspace{-1em}
\end{figure}

Unlike classic max-flow type models \cite{Dan63}, in processing-capable networks the amount of flow that arrives at a node may not be the same as the amount of flow that departs. Now, the flow conservation constraints are given by
%
\begin{align}
\sum_{j \in \mathcal{N}_i}  x^{(c)}_{ji}   + \beta^{(c)}_{i} y^{(c)}_{i} +  \lambda^{(c)}_i  =  \sum_{j \in \mathcal{N}_i}  x^{(c)}_{ij} + y^{(c)}_{i}. \label{eq:plus_flow_constraint} 
\end{align}
Equation (\ref{eq:plus_flow_constraint}) says that the data received from the other nodes in the network, plus the data received after the local processing, plus the exogenous arrivals must be equal to the traffic sent to other nodes in the network and for processing. 
Figure \ref{fig:proc_network} shows an example of a network with five IoT  nodes and a gateway. 

Finally, the decisions $\bm{x}$ and $\bm{y}$ need to respect the energy budget of each node. These must  satisfy:
\begin{align}
T_{s}  p_i(\bm{x}, \bm{y})   \le E_i &&  i \in \mathcal N \label{eq:energy-budget} 
\end{align}
where
\[
p_i(\bm{x}, \bm{y}) :=  \sum_{j\in\mathcal{N}_i}\sum_{c\in\mathcal{C}}x_{ij}^{(c)}e_{ij}^{\text{tx}} + \sum_{j\in\mathcal{N}_i}\sum_{c\in\mathcal{C}}x_{ji}^{(c)}e_{ji}^{\text{rx}}+\sum_{c\in\mathcal{C}}y_{i}^{(c)}e_{i}^{\text{pr}}
\]
and $T_{s}$ is the network life time:
\begin{definition}[Network lifetime] Time span where \emph{all} the IoT nodes are alive, \emph{i.e.}, none of them runs out of energy. 
\end{definition}
Hence, (\ref{eq:energy-budget}) must be satisfied by \emph{all} the nodes in the network. We later extend this definition to scenarios where a subset of nodes are allowed to die before the network lifetime expires.

\subsection{Utilities}
The execution of analytics at the IoT nodes and the gateway results in different performance. The latter can be related, for instance, to the precision or confidence that an image is classified using a machine learning algorithm. 
Formally, we define  function
$
\omega_{i}^{(c)}: \R \to \R$, $i \in \mathcal N, c \in \mathcal C
$
to capture the network benefit of processing an application at a node. We assume $\omega_{i}^{(c)}$ is concave and non-decreasing, and that IoT nodes may have different performance for processing the same task. This performance diversity may arise due to hardware or software differences among the IoT nodes. See  \cite[Figure 1]{RF18} for an example of how different object detection algorithms have different precisions. 

\subsection{Problem statement}
\label{sec:problem_statement}
The IoT nodes collect data that must be processed locally, at another IoT node, or at the network gateway. Ideally, we would prefer to process all the data in the gateway since this has usually an ``unlimited'' energy budget and better hardware that allows it to run more sophisticated algorithms. 
However, that might not be always possible for several reasons. First, the amount of bandwidth available between the nodes and the gateway might not be enough to transport all the data. Second, nodes are subject to energy constraints, which in turn limit the total amount of data that can be transferred. And third, the central node may not be able to cope with the processing of all the data from all the IoT nodes. 
%

Our goal is to {design a routing and processing policy ($\bm{x}, \bm{y}$) that maximizes  (i) the network lifetime and (ii) the analytics performance}. The lifetime criterion is crucial because when nodes run out of battery, the network operation is disrupted \emph{e.g.}, nodes stop collecting information, monitoring an area, among others. On the other hand, the analytics performance metric is important because our goal is not just to keep the nodes in the network alive but also to maximize the service performance. 
Sometimes these two objectives may not be conflicting since processing data in the network reduces the network congestion and so the subsequent routing energy cost. However, in-network processing incurs an energy cost as well as analytics performance degradation since  IoT nodes may run ``lighter'' algorithms that fit their hardware specifications. The overall balance depends on many parameters, such as the flow reduction due to processing, the energy cost of these tasks, and the network properties.  

In sum, we would like to find a policy $(\bm{x}, \bm{y})$ that maximizes a combined criterion of network lifetime and data analytics performance depending on {(i)} the capacity of network links; {(ii)} the nodes' processing capacity; {(iii)} the applications data rates; {(iv)} the data analytics performance ``quality''; and {(v)} the nodes' energy budget. Next, we present the mathematical formulation of the problem and introduce an algorithm that is amenable to implementation in stochastic environments. 


\section{Maximum Network Lifetime and Analytics} \label{sec:maxlife}

The IoT operation constraints determine the transmission and processing policies that are implementable, and subsequently, the set of supportable data rates. First, we define sets:
\begin{align*}
	X & : = \{ \bm{x} \mid \text{constraints } (\ref{eq:capacity_constraints}) \text{ and } (\ref{eq:interference_constraint}) \textup{ are satisfied} \} , \\
	Y & : = \{ \bm{y} \mid \text{constraint } (\ref{eq:capacity_constraints}) \textup{ is satisfied}\}.
\end{align*}
%
%
These sets are bounded polytopes (so convex sets \cite[Section 2.2.4]{BV04}) because they are the intersection of inequality constraints and all links and nodes have bounded capacity. Using these sets, we can define the capacity region of the IoT system. 
\begin{definition}[IoT network capacity region]The IoT network capacity region is the set 
\[
\Gamma (\bm{\lambda})  : = \{ \bm{x} \in X, \bm{y} \in Y \mid  \textup{constraints } (\ref{eq:plus_flow_constraint}), (\ref{eq:energy-budget}) \textup{ are satisfied}\}
\]
where $\bm{\lambda}$ is given in \eqref{eq:arrival_matrix}. 
\end{definition}
We will always assume that  $\Gamma(\bm{\lambda})$ is non-empty.
We are now in position to formulate our optimization problem.

\subsection{Maximum lifetime data analytics problem}

Recall that $T_s$ is the network lifetime. The optimization problem is given by 
\begin{align}
\begin{tabular}{lll}
$ \underset{ T_{s},  \bm{x},  \bm{y} }{\text{maximize}}$ & $\displaystyle(1-\eta) T_{s} + \eta  \sum_{i \in \mathcal{N},  c \in \mathcal{C}} \omega_{i}^{(c)}(y_{i}^{(c)}) $  &  \\
subject to & (\ref{eq:plus_flow_constraint}), (\ref{eq:energy-budget})  $\forall i \in \mathcal N$\\
&  $\bm{x} \in X, \ \bm{y}\in Y$, \  $T_{s} \ge 0$ 
\end{tabular}
\label{eq:maxlife_problem}
\end{align}
%
%
Parameter $\eta  \in [0,1]$ is used to balance how much we prioritize the network lifetime over the analytics performance metric. If $\eta = 0$, then (\ref{eq:maxlife_problem}) aims only to maximize  the network lifetime; if $\eta = 1$, the optimization only takes into account  the data analytics performance; and for any value of $\eta \in (0,1)$, it  balances the two terms in the objective. 

Problem (\ref{eq:maxlife_problem}) is non-convex because the inequality constraint (\ref{eq:energy-budget}) involves the product of variables $\bm{x}$ and $T_{s}$.\footnote{The product of two variables is generally not convex. } Moreover, variables $(\bm{x}, \bm{y})$ and $T_{s}$  are \emph{not} independent since $(\bm{x}, \bm{y})$ affect the energy consumption and so indirectly the time the network will be able to operate.  Nonetheless, and as we show in the following lemma, it is possible to transform the non-convex problem (\ref{eq:maxlife_problem}) into an ``equivalent'' convex problem. 
\begin{lemma}
\label{th:maxlifeconvex}
The optimization problem
\begin{align}
\begin{tabular}{lll}
$ \underset{ \bm{x}, \bm{y}}{\textup{minimize}}$ & 
$\displaystyle (1-\theta) \ \underset{i\in\mathcal{N}}{\max} \left\{ \frac{p_i(\bm{x}, \bm{y})}{E_i} \right\} + \theta \! \! \!\sum_{i\in\mathcal{N}, c\in\mathcal{C}} \! \! \! \! \! - \omega_{i}^{(c)}(y_{i}^{(c)} ) $  &  \\
\textup{subject to}
& \eqref{eq:plus_flow_constraint},  $\bm{x} \in X, \ \bm{y}\in Y$
\end{tabular}
\label{eq:maxlife_problem2}
\end{align}
is convex and ``equivalent'' to the non-convex problem in \eqref{eq:maxlife_problem}. It allows us to balance smoothly between maximum network lifetime and maximum data analytics performance.
\end{lemma}
\begin{IEEEproof}
Let $(T_{s}^\star, \bm{x}^\star, \bm{y}^\star)$ be a solution to (\ref{eq:maxlife_problem}). By Weierstrass' theorem \cite[Proposition 2.1.1; condition 1]{BNO03}, one can show that a solution always exists and is finite.\footnote{Note that by construction $X$ and $Y$ are bounded sets. Variable $T_s$ belongs also to a bounded set since it is nonnegative and nodes have a finite energy.}
The key part of the proof relies on showing that one of the energy constraints must be tight at the optimum. We proceed to show this by contradiction. Suppose there exists a $T' >  T^\star_s$ such that $T^\star _{s}  p_i(\bm{x}, \bm{y}) < T'  p_i(\bm{x}^\star, \bm{y}^\star) \le E_i $ for all $i \in \mathcal N$. Then, we must also have that $(1-\eta) T^\star_s < (1-\eta) T'$ (\emph{i.e.}, the objective value increases); however, this is not possible since by assumption $T_s^\star$ is an optimal value. Hence, we have that at least one constraint must be tight at the optimum, \emph{i.e.},  $T^\star _{s}  p_i(\bm{x}^\star, \bm{y}^\star) = E_i$. Next, rearrange terms in the last equation and rewrite the energy constraints as
\begin{align}
\max_{i \in \mathcal N} \left\{ {p_i(\bm{x}^\star, \bm{y}^\star)}/{E_i} \right\} = {1}/{T_{s}^\star}, \label{eq:keeqt}
\end{align}
where the $\max$ follows because the constraint must be satisfied by all the nodes. 
From (\ref{eq:keeqt}), we can see that maximizing $T_s^\star$ is equivalent to minimizing the LHS of (\ref{eq:keeqt})---which is a convex function since $p_i(\bm{x},\bm{y})$ is linear \cite[pp. 72]{BV04} and $E_i$ is a constant. Finally, considering that $\sum_{i\in\mathcal{N},c\in\mathcal{C}} - \omega_{i}^{(c)}(y_{i}^{(c)} )$ is convex, we can use scalarization with $\theta \in [0,1]$ to obtain a convex problem. 
\end{IEEEproof}
%
It is important to emphasize that problems (\ref{eq:maxlife_problem}) and (\ref{eq:maxlife_problem2}) are equivalent but {not} the same. 
Namely, if $\theta = 0$, then the optimization maximizes the network life time; if $\theta = 1$, it only takes into account the analytics performance; and if $\theta \in (0,1)$ it balances the two objectives smoothly. However, the solutions to problems (\ref{eq:maxlife_problem}) and (\ref{eq:maxlife_problem2}) do \emph{not} need to be the same for $\theta = \eta$ when $\theta \in (0,1)$ since the network lifetime term in (\ref{eq:maxlife_problem}) is linear, whereas in (\ref{eq:maxlife_problem2}), the term must be regarded as $1/T_s$. 
%
As we will show in detail in Section \ref{sec:evaluation}, parameter $\theta$ has a huge impact on the system's performance.

\subsection{Generalized maximum network lifetime problem} 
Instead of considering the lifetime of individual nodes, we can consider the lifetime of groups of nodes $\mathcal{M}_k \subset \mathcal N$, $k \in \{1,2,\ldots,K\}$. For instance, the type of data collected by a subset of nodes may be more important to the global system objective than the data collected by another subset of nodes. To consider groups of nodes in the optimization problem, we can replace the network lifetime term in the objective of problem (\ref{eq:maxlife_problem2}) with 
\[
 \sum_{k=1}^K \pi_k\max_{i\in\mathcal{M}_k} \{ {p_i(\bm{x}, \bm{y})}/{E_i} \}.
\]
Parameter $\pi_k \ge 0$ is used to emphasize how much we would like a subset of nodes to ``remain'' alive with respect to another subset.
Note that the formulation in (\ref{eq:maxlife_problem2}) is a special case where $| \mathcal M_k| = 1$ for all $k \in \{1,\dots,K\}$ and $\cup_{k=1}^K \mathcal M_k= \mathcal N$. Hereafter, and to streamline exposition, we will use the formulation in (\ref{eq:maxlife_problem2}) but our results apply to more general cases directly. 

\subsection{Practical limitations}
The resulting optimization problem is readily solvable by convex solvers such as SCS \cite{DCP+17}. However, for this, one needs to know all the system parameters, which is rarely the case in practical scenarios. For example, the average capacity of a link connecting two IoT nodes is usually not known. Furthermore, there are generally instantaneous constraints, such as the level of noise in the system or interference, which affect the decisions that the IoT nodes can make. 
In the next section, we present a dynamic algorithm that learns and adapts to the instantaneous network/system characteristics. 

\section{Dynamic MLIA Algorithm}\label{sec:dynamic_problem}

In this section, we present a dynamic algorithm that solves (\ref{eq:maxlife_problem2}) and does not require previous knowledge on (i) the average arrival rate of the data analytics; (ii) the capacity of the network links $\bm{\mu}$ in each time instance; and (iii) the average performance or reward obtained from carrying out data analytics at the IoT nodes. 

\subsection{Dynamic problem formulation}

We divide the time in slots $t \in \N$ of normalized duration and  parameterize the variables in the static model \eqref{eq:maxlife_problem2} with $[t]$ to indicate their value in a time slot. For instance, $\lambda_{i}^{(c)}[t]$ indicates the new arrivals of application $c \in \mathcal{C}$ at node $i$ and time $t \in \N$, and $x^{(c)}_{ij}[t]$ is the amount of commodity transmitted over link $(i,j) \in \mathcal E$.   
We also need to capture the network constraints of the dynamic problem. 
The instantaneous link capacity constraints are
\begin{align}
\sum_{c \in \mathcal C} x^{(c)}_{ij} [t] \le \mu_{ij} [t], \qquad  \sum_{c \in \mathcal C} \gamma_{i}^{(c)} y^{(c)}_{i} [t] \le \rho_i [t]  \label{eq:capacity_constraint_instantaneous}
\end{align}
$\forall (i,j) \in \mathcal E,  i \in \mathcal N$. The (binary) interference constraints are 
\begin{align}
\!\!\!\!\textstyle \mathbb{I} \left( \sum_{c\in\mathcal{C}}x_{ij}^{(c)}[t] \right) 
+ \sum_{(k,m)\in I(i,j) } \mathbb I \left({\sum_{c\in\mathcal{C}}x_{km}^{(c)}}[t] \right) \leq 1, \label{eq:interference_constraint_instantaneous}
\end{align}
for all $(i,j)\in\mathcal{E}$ where $\mathbb I$ is the indicator function, \emph{i.e.,} $\mathbb I(x) = 1$ if $x > 0$ and $\mathbb I(x) = 0$ otherwise. Hence, the binary interference constraints only allow one node to transmit in the interference range regardless of the transmission rate. 
We are now in position to define sets
\begin{align*}
	X[t] & : = \{ \bm{x} \mid \text{constraints } (\ref{eq:capacity_constraint_instantaneous}) \text{ and } (\ref{eq:interference_constraint_instantaneous}) \textup{ are satisfied} \} , \\
	Y[t] & : = \{ \bm{y} \mid \text{constraint } (\ref{eq:capacity_constraint_instantaneous}) \textup{ is satisfied}\},
\end{align*}
which contain the admissible routing/processing policies that can be implemented in the system in each time slot. 
%
Note that $X[t]$ may not be convex since (\ref{eq:interference_constraint_instantaneous}) is not convex. 

The operation of the network consists of selecting actions (or values) from the instantaneous action sets $X[t]$ and $Y[t]$ while taking into account the properties of the underlying convex problem. 
We explain next how this can be achieved.

\subsection{Algorithm overview}

The key idea for solving (\ref{eq:maxlife_problem2}) in the dynamic setting is to relax the flow conservation constraints and formulate the Lagrange dual problem. The Lagrangian is
\begin{align*}
& L(\bm{x}, \bm{y}, \bm{\nu} )  =  
(1-\theta) \ \underset{i\in\mathcal{N}}{\max} \left\{ \frac{p_i(\bm{x}, \bm{y})}{E_i} \right\} + \theta \! \! \!\sum_{i\in\mathcal{N},c\in\mathcal{C}} \! \! \! \! \! - \omega_{i}^{(c)}(y_{i}^{(c)} ) \\
& + \sum_{i \in \mathcal N} \sum_{c \in \mathcal C}\nu_i^{(c)} \left( \sum_{j\in\mathcal{N}_i} (x_{ji}^{(c)} - x_{ij}^{(c)}) + (\beta_i^{(c)} - 1)y_{i}^{(c)} + \lambda_i^{(c)} \right)
\end{align*}
where $\nu_i^{(c)} \in \R$ is the Lagrange multiplier associated to each of the flow conservation constraints, and we define $\bm{\nu} = (\nu^{(c)}: i \in \mathcal{N}, c \in \mathcal{C})$.
The Lagrange dual function is defined as 
\begin{align}
h(\bm{\nu}) := \min_{\bm{x} \in X, \bm{y} \in Y} L(\bm{x},\bm{y}, \bm{\nu}), \label{eq:lagrange_dual}
\end{align}
and recall that it is concave \cite[Section 5.2]{BV04}. Hence, it can be maximized with the subgradient method applied to the Lagrange dual problem.\footnote{We use subgradient instead of gradient because the Lagrange dual function may not be differentiable.} Specifically, with update:
\begin{align}
\textstyle \nu_i^{(c)}[t+1] = &  \textstyle\nu_i^{(c)}[t]  +  \alpha \Big( \sum_{j\in\mathcal{N}_i} (x_{ji}^{(c)}[t] - x_{ij}^{(c)}[t])  \label{eq:dual_update}\\
&\textstyle  +  (\beta_i^{(c)} - 1)y_{i}^{(c)}[t] +  \lambda_i^{(c)} \Big) \qquad \forall c \in \mathcal C, i \in \mathcal N , \notag
\end{align}
where the term in parenthesis is a subgradient of $h$ at $\bm{\nu}[t]$, 
\begin{align}
(\bm{x}[t], \bm{y}[t]) \in \argmin_{\bm{x} \in X, \bm{y} \in Y} L(\bm{x},\bm{y},\bm{\nu}),\label{eq:compute_gradient_partial}
\end{align}
and $\alpha >0$ the step size or parameter that controls the accuracy of solution in the optimization.
Recall that the solution of the primal and dual problem coincides when strong duality holds \cite[Section 5.2.3]{BV04}; which is the case in our problem since the constraints in (\ref{eq:maxlife_problem2}) are linear and the network capacity or feasible set $\Gamma(\bm{\lambda})$ is non-empty by assumption.

Next, we proceed to explain how to include constraints (\ref{eq:capacity_constraint_instantaneous})-(\ref{eq:interference_constraint_instantaneous}) so that our dynamic algorithm can be implemented in a real system. We do this incrementally from the static problem.
\subsubsection{Unknown arrivals} Suppose now that $X[t] = X$ and $Y[t] = Y$ for all $t \in \N$ (the sets of actions do not change over time). 
We want to relax the fact that the data analytics arrival  $\bm{\lambda}$ is not known a priori.
For this, we need to solve the Lagrange dual problem (\emph{i.e.}, maximize $h(\bm{\nu})$) with the subgradient method replacing  $\lambda_i^{(c)}$ with the random variable $\lambda_i^{(c)}[t]$ in the dual update (\ref{eq:dual_update}). This change amounts to making the subgradient of $h(\bm{\nu})$ stochastic, or equivalently, to using the stochastic dual subgradient method to solve the underlying convex problem \eqref{eq:maxlife_problem2}. For the algorithm to converge we need to make the following assumption.
\begin{assumption}
\label{as:iid_arrivals}
$\{ \lambda_{i}^{(c)} [t] \}$ is an i.i.d.\ process with expected value $\lambda_{i}^{(c)}$ for all $i \in \mathcal N$, $c \in \mathcal C$, $t \in \N$ and $\lambda^{(c)}_i[t]$ is uniformly bounded for all $t \in \N$. 
\end{assumption}

\subsubsection{Time-varying sets}
We can incorporate time-varying sets in our algorithm by replacing $X$ and $Y$ with $X[t]$ and $Y[t]$ in update (\ref{eq:compute_gradient_partial}). In order to solve the underlying problem \eqref{eq:maxlife_problem2}, we need to make the following assumption, which can be regarded as if we used stochastic subgradients in update \eqref{eq:dual_update}.
\begin{assumption}
\label{as:iid_sets}
$\E(X[t]) = X$ and $\E(Y[t]) = Y$ for all $t \in \N$ and $X[t]$ and $Y[t]$ are bounded sets. Here, the expectations are defined using (Minkowski) set addition \cite[pp. 32]{GNT06}.
\end{assumption}
%


\subsubsection{Noisy utility function} 
We may not have access to the utility function that measures the \emph{exact} reward derived from processing analytics locally. To capture this, we define 
$
\tilde \omega^{(c)}_i: \R \to \R, i \in \mathcal N, c \in \mathcal C
$
to be an estimate of utility function $w_i^{(c)}$. Also, let $\tilde L(\bm{x}, \bm{y}, \bm{\nu} )$ be the Lagrangian where $\tilde \omega^{(c)}_i$ is used instead of $ \omega^{(c)}_i$, and replace the Lagrangian in (\ref{eq:compute_gradient_partial}) with  $\tilde L(\bm{x}, \bm{y}, \bm{\nu} )$.
Technically, minimizing $\tilde L(\bm{x}, \bm{y}, \bm{\nu} )$ instead of $L(\bm{x}, \bm{y}, \bm{\nu} )$ corresponds to computing (sub)gradients of the Lagrange dual function approximately, \emph{i.e.}, having $\epsilon$-subgradient \cite[pp. 625]{Ber99}. The convergence depends on the mild assumption that the errors are bounded. 
\begin{assumption}
\label{th:boundedferror}
The maximum error between  $ \omega^{(c)}_i$ and  $\tilde \omega^{(c)}_i$ is bounded, \emph{i.e.}, $\max_{y \le \rho_i} | \omega^{(c)}_i (y) - \tilde \omega^{(c)}_i (y)| : = \xi_i^{(c)} < \infty$.
\end{assumption}


\subsection{Algorithm \& convergence} 

Algorithm \ref{al:algorithm_v1} consists of three steps. The first one is to obtain the network state or set of possible actions, \emph{i.e.}, $X[t]$ and $Y[t]$. We assume these are obtained by the IoT nodes and that the information is transmitted to the network gateway where the routing decisions are made. The second step is to minimize the Lagrangian using sets $X[t]$ and $Y[t]$. The complexity of this step depends on the number of elements in the sets. When these are discrete and contain few elements, the minimization can be carried out by exhaustive search, and if $X[t]$ and $Y[t]$ are convex by standard convex optimization techniques \cite{BV04, BNO03}. The third and final step is to update the dual variables, \emph{i.e.}, carry out the (stochastic) subgradient update with $\epsilon$-subgradients. Parameter $\alpha > 0$ is the step size that controls the accuracy/speed tradeoff of the algorithm \cite{BNO03}. 

We establish the convergence of the algorithm with respect to an optimal policy implemented in a random fashion. We need to  make the following definitions.
\begin{definition}[Average policy space]
\[
X := \lim_{T \to \infty }\frac{1}{T}\bigoplus_{t=1}^T X[t], \quad  Y := \lim_{T \to \infty }\frac{1}{T}\bigoplus_{t=1}^T Y[t]
\]
where $\oplus$ denotes the (Minkowski) set addition \cite[pp. 32]{GNT06}.
\end{definition}
\begin{definition}[Optimal policy] An optimal policy is a pair $(x^\star,y^\star) \in (X, Y)$ that solves \eqref{eq:maxlife_problem2}. By construction we have 
\begin{align*}
x^\star  = \lim_{T \to \infty } \frac{1}{T} \sum_{t = 1}^T x^\star[t] , \quad y^\star  = \lim_{T \to \infty } \frac{1}{T} \sum_{t = 1}^T y^\star[t] 
\end{align*}
where $x^\star [t] \in X[t]$ and $y^\star [t] \in Y[t]$ for all $t \in \N$. 
\end{definition}
%


\begin{theorem}
Let $f$ be the objective function in the optimization problem \eqref{eq:maxlife_problem2}. Suppose Assumptions \ref{as:iid_arrivals}, \ref{as:iid_sets} and \ref{th:boundedferror} are satisfied. Algorithm \ref{al:algorithm_v1} ensures that
\begin{align*}
& \textup{(i)} \quad   \frac{1}{T} \sum_{t=1}^T  \E \left( f (  x[t] , y[t] ) -  f (  x^\star[t] , y^\star [t] )  \right)   \le \alpha \epsilon_1 +  \epsilon_2, \\
& \textup{(ii)} \lim_{T \to \infty} \E \left(
(\bar x_{ji}^{(c)} - \bar x_{ij}^{(c)}) - (\beta_i^{(c)} - 1) \bar y_{i}^{(c)} + \lambda_i^{(c)}[t]
\right) = 0,  
 \end{align*}
$\forall i \in \mathcal N$, $c \in \mathcal C$ where  $\bm{\bar x} := \frac{1}{T} \sum_{t=1}^T  {\bar x} [t]$,  $\bm{\bar y} = \frac{1}{T} \sum_{t=1}^T  \bm{y} [t]$; $\epsilon_1$ is a bounded constant related to the variance of the arrival process and sets $X[t]$ and $Y[t]$, and $\epsilon_2$ a constant such that $| \xi_i^{(c)} |_2 \le \epsilon_2$. \label{th:main_theorem}
\end{theorem}
Theorem \ref{th:main_theorem} establishes an upper bound on the objective function of our policy $(\bm{x}[t], \bm{y}[t])$ compared to an optimal policy $(\bm{x}^\star, \bm{y}^\star)$ implemented in a ``randomized'' fashion---that is, the policy that we would implement if we knew in advance all the parameters in the system. Note that the bound on the optimality gap depends on $\epsilon_1$ and $\epsilon_2$. The first term is related to the variation of the arrivals and link capacities, and the second on how good the estimate of the reward functions are. Importantly, note that the first term depends on  the step size $\alpha$, which means that we can make it arbitrarily small. 
Theorem \ref{th:main_theorem} establishes also that the flow conservation constraints are satisfied on expectation as $T \to \infty$, which means that we will recover, asymptotically, a policy in the network capacity region $\Gamma (\bm{\lambda})$. 

\subsection{Practical aspects}


\subsubsection{Non-asymptotic analysis} Differently from previous stochastic network optimization analyses, \emph{e.g.}, \cite{ES07+, Nee10}, our results do \emph{not} compare the performance of the average policy. Furthermore, and unlike these previous works that give optimality bounds only asymptotically (\emph{i.e.}, as $T \to \infty$), here we provide guarantees on how the policy performs on expectation per time slot. This is very important for this problem because IoT nodes have a \emph{finite} energy budget that may not allow them to reach a ``steady'' state---this will be illustrated in the experiments in Section \ref{sec:performance}.

\subsubsection{Distributed execution} MLIA describes a centralized algorithm where the IoT gateway collects the nodes' parameters to solve \eqref{eq:compute_subgradient}. However, the only necessary central calculation is that for devising an eligible link activation schedule (based on the interference constraints). If such a schedule is already given or if the links are orthogonal (or, if any other interference control scheme is used) then each node can independently optimize its actions. Namely, each node $i$ can calculate separately the quantity $\partial \tilde L(\bm{x}, \bm{y}, \bm{\nu})/\partial \bm{x}_{ij}$, $\forall j\in\mathcal{N}_i$, and similarly for $y_{ij}$ variables; and then each pair of 1-hop neighbors exchange the respective dual variables. This is a standard approach for enabling a decentralized implementation of such protocols, see \cite{GNT06} for a survey, and applies directly to MLIA.

\begin{algorithm}[t]
\caption{Maximum Lifetime IoT Analytics (MLIA)}
\begin{algorithmic}[0]
	\STATE {\bfseries Parameters:}  step size $\alpha \ge 0$
		\STATE Initialize $t = 1$; $\nu_{i}^{(c)}[t]= 0, \ \text{for all} \ c \in \mathcal{C}, i \in \mathcal{N}$. 
	\STATE{\textbf{In each time slot } $t=1,2,\dots$}
	\begin{itemize}
		 \item  [1)]  \underline{\smash{Obtain network state:}} the network gateway collects the network connectivity information and computes sets $X[t]$ and $Y[t]$.
		\item  [2)] \underline{\smash{Compute action}}: the network gateway obtains		\begin{align}
		\big(\bm{x}[t], \bm{y}[t]\big) \in \! \! \! \! \argmin_{ \substack{\bm{x} \in X[t], \bm{y}\in  Y[t]} } \tilde L(\bm{x}, \bm{y}, \alpha \bm{\nu}[t]) \label{eq:compute_subgradient}
		\end{align}
		and broadcast the solution to the IoT nodes. 
		
		\item  [3)] \underline{\smash{Update the dual variables:}} for all ${\nu}_{i}^{(c)}$, $i \in \mathcal N$, $c \in \mathcal C$ perform update (\ref{eq:dual_update}) with  $\lambda_i^{(c)}[t]$ instead of $\lambda_i^{(c)}$. 
		\end{itemize}
		\textbf{end loop}
\end{algorithmic}
\label{al:algorithm_v1}
\end{algorithm}

\section{Experiments and Evaluation}
\label{sec:evaluation}

We evaluate the performance of our approach numerically using real hardware and application parameters. We investigate three points. First, how parameter $\theta$ and the network connectivity affect the network lifetime and the analytics performance in the {static} or offline problem \eqref{eq:maxlife_problem2}. Second, the convergence of the \emph{dynamic} algorithm to the solution of the static problem depending on $\alpha$. And third, how the proposed algorithm compares to two benchmark algorithms. We fix the network size in the experiments to $20$ nodes (including the gateway), but similar results are obtained with different network sizes.  

\subsection{Experiments setup}
\subsubsection{Network} In all simulations, we use random geometric graphs (RGG) to model an IoT network---this is common practice in wireless networks \cite{HAB+09}.
Recall that in RGGs nodes are placed uniformly at random on an area of $1 \times 1$ normalized units, and that their connectivity depends on the \emph{radius} or \emph{distance} each node covers.
For simplicity, we assume that \emph{all} links have an average capacity of $24$ \si{\mega\bits}. 

\subsubsection{Nodes} The IoT nodes are Raspberry Pi(es) 3B  equipped with an ARMv7 CPU, 1 \si{\giga \byte} RAM and a 802.11.b/g/n network card. According to our measurements, the power required to transmit and receive data is $0.4$\,\si{\watt} (with small variations depending on the channel quality), and $2.1$\,\si{\watt} for processing at full power. We assume that the energy spent to collect data and in idle mode is negligible.\footnote{That is, IoT nodes can switch to low-power consumption mode.} All IoT nodes have batteries with a capacity of $2500$ \si{\joule};\footnote{The batteries have a small capacity to keep simulations short. An IoT node can easily be equipped with a battery that has one hundred times more energy.} the gateway is connected to a constant energy source and so does not have energy constraints.  

\subsubsection{Application} This consists of analyzing video streams of rate $1$ \si{\frame/\second} with the object detection algorithm YOLO \cite{RF18}. Frames have size $0.5$ \si{\mega\byte} and the outcome of the processing  (a collection of bounding boxes and tags that indicate where the objects are in a frame; see the video in \cite{yolo_web} for an example) can be represented with at most $0.5$ \si{\kilo\byte}. Hence, $\beta = 0.5 \si{\mega\byte} /  0.5 \si{\kilo\byte} = 10^{-3}$. 
We  consider that an IoT node is a ``source'' node (receives a video stream) with probability $1/2$.  According to our measurements, the IoT nodes and gateway (a desktop with a GPU) can process a frame in $3$ \si{\second} and $100$ \si{\milli\second} respectively. The analytics performance metric is given by the mean Average Precision (mAP \cite{mAP}) that YOLO can detect objects correctly in a frame  This is $33.1$ (YOLOv3-tiny) for an IoT node and $57.9$ (YOLOv3-608)  for the gateway \cite{yolo_web}.

\subsection{Evaluation}



\begin{figure}
\centering
{\resizebox{0.9\columnwidth}{!}{\input{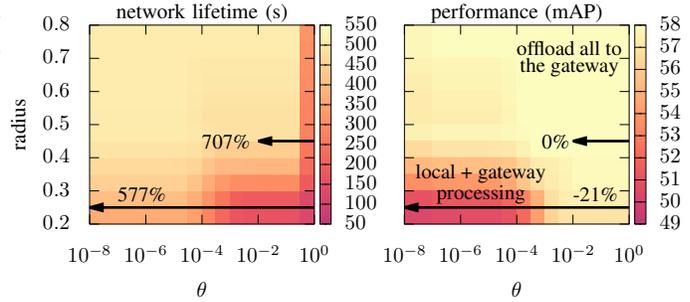}}}
\vspace{1.5em}
\caption{Illustrating the sensitivity of the solution depending on parameter $\theta$ and the network radius.}
\label{fig:sensitivity_exp}
\end{figure}

\subsubsection{Sensitivity analysis}

Figure \ref{fig:sensitivity_exp} shows the network lifetime (left) and data analytics performance (right) for a range of values $\theta$ and nodes' radius. Recall that the radius affects the structure of the graph (\emph{e.g.}, number of links, shortest path to the gateway). The results are the average of 50 different networks generated as described in the previous section.  Observe from the figure that independently of the radius, the network lifetime increases monotonically as $\theta \to 0$.  Nonetheless, the radius plays an important role: when the network connectivity is high,\footnote{Nodes are on average less than two hops away from the gateway.} the best strategy is to offload all the processing to the gateway.\footnote{That is because processing analytics at the nodes locally is more expensive than transmitting them, and the processing reward at the gateway is larger than the reward obtained as a result of processing the analytics at the nodes.}  Otherwise, the solution balances between local and gateway processing. Also, observe that in this specific example selecting $\theta = 1$ is generally not the ``best'' choice. 
When the radius is equal to $0.25$, if we change from $\theta = 1$ to $\theta = 10^{-8}$ we can increase the network lifetime by $577\%$ at the expense of just $-21 \%$ in the data analytics performance (see Figure \ref{fig:sensitivity_exp}). Similarly, when the radius is equal to $0.45$ (the network is dense), by changing $\theta$ from $1$ to $10^{-8}$ the network lifetime increases by $707\%$ without affecting the analytics ($0 \%$ change). \textbf{Conclusion:} the data  analytics performance and network lifetime terms in the objective are very sensitive to parameter $\theta$. Also, these may not necessarily conflicting: it is possible to obtain both, good analytics performance and network lifetime, by setting $\theta$ properly.

\begin{figure}
\centering
{\resizebox{0.9\columnwidth}{!}{\input{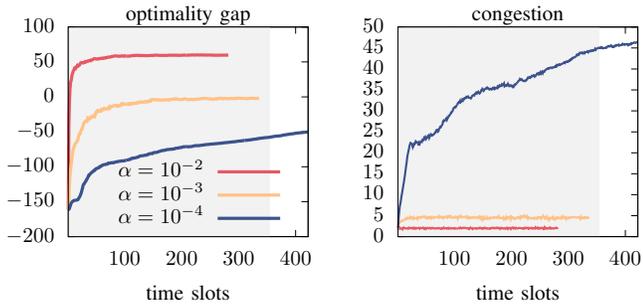}}}\\
\vspace{1.5em}
\caption{Illustrating the relationship between the optimality gap and the congestion in the network depending on the step size $\alpha$.}
\label{fig:dynamicplot_exp}
\vspace{-1.5em}
\end{figure}

\subsubsection{Performance of the dynamic algorithm}
\label{sec:performance}
We now investigate how Algorithm \ref{al:algorithm_v1} solves the static problem for a network with radius equal to $0.25$. Parameter $\theta$ is fixed to $10^{-5}$. The video frames arrive in the IoT nodes following a Poisson process. To capture the errors on the prediction function, we add noise of $20 \%$. 
We run Algorithm \ref{al:algorithm_v1} with $\alpha \in  \{10^{-2}, 10^{-3}, 10^{-4}\}$ and show the results in Figure \ref{fig:dynamicplot_exp}. The results are the average of 50 samples. On the left, we have the normalized optimality gap\footnote{The optimality gap divided by $\theta$.} per iteration, and on the right, the system's congestion or sum of all the Lagrange multipliers $\nu^{(c)}_i$. The gray area indicates the lifetime of the system computed in the static problem. 

Observe that the plots have different lengths (\emph{i.e.}, different network lifetimes) and convergence behavior. Specifically, with $\alpha = 10^{-2}$, the transient phase is short,\footnote{Time required to converge to a ball or steady value around the optimum.} the optimality gap is large, and the network lifetime shorter than in the static problem. With $\alpha = 10^{-3}$, the transient phase is slightly longer, but we obtain instead a much smaller optimality gap, and the network lifetime matches nearly the one of the static problem. Finally, with  $\alpha = 10^{-4}$, the transient phase is so long that the optimality gap is always negative (\emph{i.e.}, better than the \emph{static} optimum) and the network lifetime longer than in the static problem. However, note that this is possible because congestion keeps building up in the system, \emph{i.e.}, data analytics get accumulated in the nodes without being processed neither transmitted. \textbf{Conclusion:} parameter $\alpha$ controls not only the optimality gap but also the duration of the transient phase. Selecting $\alpha$ too small may result in generating congestion and not routing/processing data analytics. 

\subsubsection{Comparison with other algorithms} 
\label{sec:compare}
\begin{figure}
\centering
{\resizebox{0.4\columnwidth}{!}{\input{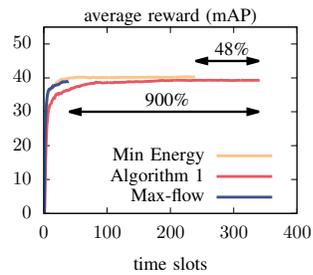}}}\\
\vspace{1.5em}
\caption{Illustrating how the Algorithm \ref{al:algorithm_v1} compares to a min-energy and max-flow type algorithms. }
\label{fig:compare_exp}
\vspace{-1em}
\end{figure}

Given that this paper introduces a new problem, there are no other algorithms with which to  compare. Hence, we use two intuitive alternatives: (i) a max-flow type algorithm \cite{Dan63} that makes routing/processing decisions based only on the network congestion (\emph{i.e.}, the Lagrange multipliers); and (ii) an algorithm that balances analytics performance and energy consumption (instead of network lifetime). Specifically, the first term in the objective in problem (\ref{eq:maxlife_problem}) is replaced with $\sum_{i \in \mathcal N} p_i(\bm{x}, \bm{y})$.\footnote{We assume  that averages $\bm{\lambda}$ and $\bm{\mu}$ are known in the min-energy algorithm.}  Also, to compare the algorithms fairly, we fix the reward for processing analytics to $40$ mAP and evaluate their performance in terms of network lifetime gain (we do this by selecting $\eta$ and $\theta$ accordingly). 
Figure \ref{fig:compare_exp} shows the average data analytics reward for 50 different realizations (time-varying arrivals and link capacities). Parameter $\alpha$ is selected equal to $10^{-3}$. Observe that with the proposed algorithm, we obtain $48\%$ longer lifetime than with the algorithm that minimizes only the total energy consumption (min-energy), and a $900\%$ gain compared to the algorithm that only considers the network congestion (max-flow). \textbf{Conclusion:} minimizing the total energy consumption does not necessarily maximize the network lifetime. Algorithms that do not take into account the energy constraints can degrade the lifetime of the network significantly. 

\section{Conclusions}

We have studied a crucial new problem in emerging IoT networks: how to allocate bandwidth and computation resources to data analytics tasks while considering the time the network can operate. The algorithm proposed can (i) balance between maximizing the network lifetime and the grade in which analytics are carried out; and (ii) operate without knowledge of the traffic arrivals or channel statistics. Our simulations seeded by real IoT device energy measurements, show that the network connectivity 
plays a crucial role in network lifetime maximization, 
that the algorithm can obtain both maximum network lifetime and maximum data analytics performance 
in addition to maximizing the joint objective, 
and that the algorithm increases the network lifetime by approximately 50\% compared 
to an algorithm that minimizes the total energy consumption in the network.

\bibliographystyle{IEEEtran}
\bibliography{references}


\section{Appendix}
\label{section:analysis}

\subsection{Preliminaries}
To streamline exposition and due to space constraints, we assume in the analysis there is only one commodity. The extension to multiple commodities is nonetheless straightforward. Now, we write the Lagrangian as
$
 L({x}, {y}, {\nu} )= f(x, y) +  \nu^\top   (g(x,y) + \lambda),
$
where $f(x,y)$ is the objective function in (\ref{eq:maxlife_problem2}) and $g(x,y) = (g_1(x,y), \dots, g_n(x,y))$ with  $g_i(x,y) =  (\beta_{i}-1) y_{i}  + \sum_{j \in \mathcal{N}_i}  (x_{ji} - x_{ij})$, $i \in \{1,\dots,n\}$. That is, $g_i(x,y) + \lambda_i$ is the  flow conservation constraint of node $i \in \mathcal N$ and $g(x,y)$ a collection of $n$ flow conservation constraints.  Note we do not use bold notation and that $\nu^\top   (g(x,y) + \lambda)$ is the inner product of vectors $\nu$ and  $(g(x,y) + \lambda)$.

%

%

%


\subsection{Proof of Theorem \ref{th:main_theorem}}

We have the dual (sub)gradient update 
$
\nu[t+1] =\nu[t]+\alpha (g({x}[t], {y}[t]) + \lambda) 
$
where  ${x}[t] \in X[t]$ and ${y}[t] \in Y[t]$ are obtained by minimizing the Lagrangian for a fixed $\nu[t]$.
We start by showing an upper bound on the objective function. 

\begin{lemma}[Objective upper bound] 
\label{th:objective_upper_bound}It holds
\begin{align}
& \textstyle \frac{1}{T} \sum_{t= 1}^T (f(x{[t]}, y[t]) - f(x^\star{[t]},  y^\star[t]) )   \label{eq:twobounds} \\
& \textstyle \le - \frac{1}{T}\sum_{t= 1}^T (\nu[t])^\top (g(x[t],y[t]) + \lambda[t]) \notag \\
& \textstyle \quad +  \frac{1}{T} \sum_{t= 1}^T (\nu[t])^\top (g(x^\star[t], y^\star[t]) + \lambda[t]) \notag + \epsilon_2   \notag
\end{align}
\end{lemma}
\begin{IEEEproof}
Since $x[t]$ and $y[t]$ are selected from $X[t]$ and $Y[t]$ to minimize the Lagrangian, we can write 
$
\tilde f(x{[t]}, y[t]) + (\nu[t])^\top (g(x[t],y[t]) + \lambda[t]) 
  =   f(x{[t]}, y[t]) + \xi(x[t],y[t])  + (\nu[t])^\top (g(x[t],y[t]) + \lambda[t]) 
  \le  f(x^\star{[t]}, y^\star[t]) + \xi(x^\star[t], y^\star[t]) 
  + (\nu[t])^\top (g(x^\star [t], y^\star [t]) + \lambda[t])
$.
Rearranging terms, summing from $t = 1, \dots, T$, dividing by $T$,  and using the fact that $| \xi(x[t],y[t])| \le \epsilon_2$ by assumption for all $(x[t],y[t]) \in (X[t],Y[t])$ yields the stated result. 
\end{IEEEproof}

Next, we bound the first two terms in the RHS of (\ref{eq:twobounds}). 

\begin{lemma} \label{th:up1}
The first term in the RHS of (\ref{eq:twobounds}) is upper bounded by $\frac{\alpha}{2T} \sum_{t = 1}^T \sigma[t] $ where $\sigma[t] := \| g(x[t],y[t]) + \lambda[t] \|_2^2$. 
\end{lemma}
\begin{IEEEproof}
Let $\theta \in \R^n$. From the standard dual subgradient update, we have that
$
\| \nu[t+1] -\theta \|_2^2 
 = \|  \nu[t] + \alpha (g(x[t],y[t]) + \lambda[t])   - \theta\|_2^2 
 = \|  \nu[t] - \theta \|_2^2 + \alpha^2 \| g(x[t],y[t]) + \lambda[t] \|_2^2   
   + 2 \alpha ( \nu[t] - \theta)^\top (g(x[t],y[t]) + \lambda[t])  
 \le \|  \nu[t] - \theta \|_2^2 + \alpha^2 \sigma[t] + 2 \alpha ( \nu[t] - \theta)^\top (g(x[t],y[t]) + \lambda[t]).
 $
%
Parameter $\sigma [t]$ is bounded since $X[t]$, $Y[t]$, $\lambda[t]$ are bounded for all $t \in \Z_+$.
Next, rearrange terms, apply the expansion recursively for $t = 1,\dots, T$ to obtain $-  2 \alpha \sum_{t= 1}^T  (\nu[t] - \theta)^\top (g(x[t],y[t]) + \lambda[t]) \le \alpha^2 \sum_{t= 1}^T \sigma[t] + \| \nu[1] -\theta \|_2^2 - \| \nu[t] - \theta \|_2^2$. 
%
Drop the third term in the RHS of the last equation since it is nonnpositive, let $\theta = 0$ and $\nu[1] = 0$, and divide across by $2 \alpha T $ to obtain the stated bound. 
\end{IEEEproof}

\begin{lemma}
\label{th:up2}
The second term in the RHS of (\ref{eq:twobounds}) is equal to  zero on expectation. 
\end{lemma}

\begin{IEEEproof}
Take expectations with respect to random variable $X[t]$, $Y[t]$, $\lambda[t]$, and write 
$
\E ( \frac{1}{T} \sum_{t= 1}^T (\nu[t])^\top (g(x^\star[t],y^\star[t]) + \lambda[t])) 
\stackrel{\text{(a)}}{=} \frac{1}{T} \sum_{t= 1}^T (\nu[t])^\top \E (g(x^\star[t],y^\star[t]) + \lambda[t])   \stackrel{\text{(b)}}{=} 0 
$
where (a) follows from the linearity of the expectation, and (b) since $\E (g(x^\star[t],y^\star[t]) + \lambda[t]) = \E (g(x^\star[t],y^\star[t])) + \E(\lambda[t]) =  g(x^\star,y^\star) + \lambda  = 0$ (\emph{i.e.}, at the optimum the problem must be feasible). Note that this is always case since random variables $X[t]$, $Y[t]$, $\lambda[t]$ do not depend on $\nu[t]$ for all $t \in \Z_+$. 
\end{IEEEproof}

We are now in position to prove claim (i). Take expectations in the bound obtain in Lemma \ref{th:objective_upper_bound}. The first term can be upper bounded by Lemma \ref{th:up1} and the second with Lemma \ref{th:up2}. By letting $\E(\sigma[t]) \le 2 \epsilon_1$ for all $t \in \Z_+$ the stated result follows.


\begin{lemma}[Feasibility]
\label{th:feasibility}
 $\E(g( \bar x,  \bar y) + \lambda) = 0$ converges to a feasible point asymptotically as $T \to \infty$. \end{lemma}
\begin{IEEEproof}[Sketch]
Recall $\nu[t+1]  =  \nu[t] + \alpha(g(x[t],y[t]) + \lambda[t]))$. 
Rearrange terms and apply the iteration recursively to obtain
$
\nu[t+1] - \nu[1] = \sum_{t=1}^T \alpha (g(x[t],y[t]) + \lambda[t])).
$
Dividing by $T$ and using the fact that $\nu[1]= 0$ we have
$
\frac{1}{T} \sum_{t=1}^T  g(x[t],y[t]) + \lambda[t]  = g( \bar x,  \bar y) +  \bar \lambda  = {\nu[t]}/({\alpha T}) \notag 
$
which implies that $(g( \bar x,  \bar y) + \bar \lambda) \to 0$ as $T \to \infty$ if $\nu[t]$ is bounded. The latter will be the case, on expectation, when $\nu[t]$ is a nonnegative process (nodes do not pre-serve/process data) and the flow conservation constraints are not tight: there exists a $\chi \succ 0$ and $(\hat x,\hat y)$ such that $g(\hat x,\hat y) + \lambda + \chi = 0$ holds. 
%
\end{IEEEproof}

\end{document}